\documentstyle[11pt,paspconf]{article}
\newcommand{\ea}{{\it et al.}}

\newcommand{\beq}{\begin{equation}}
\newcommand{\eeq}{\end{equation}}	
\newcommand{\msun}{\hbox{$ {\rm M}_{\odot}$}}

\begin{document}

\title{The Evolution of Procyon A}
\author{Brian Chaboyer\altaffilmark{1}}
\affil{Steward Observatory, University of Arizona, Tucson, AZ 85721}

\author{P. Demarque }
\affil{Department of Astronomy, and Center for Solar and
Space Research, Yale University, Box 208101, New Haven, CT 06520-8101}
\author{D.B. Guenther}
\affil{Department of Astronomy and Physics, Saint Mary's University, 
Halifax, Nova Scotia}

\begin{abstract}
A grid of stellar evolution models for Procyon A has been calculated.
These models include the best physics available to us (including the
latest opacities and equation of state) and are based on the revised
astrometric mass of Girard {\it et al.}\ (1996).  The long standing discrepancy
between the evolutionary mass and the astrometric mass is now resolved,
a result of the newly determined astrometric mass.  Models were
calculated with helium diffusion and with the combined effects of
helium and heavy element diffusion.  Oscillation frequencies for
$\ell=0,1,2$ and 3 $p$-modes (and $g$-modes) were calculated for these
models.  The predicted $p$-mode frequencies are relatively unaffected
by heavy element diffusion and convective core overshoot.  The
inclusion of a modest stellar wind which effectively suppresses the
helium diffusion in the surface layers has a modest effect on the
$p$-mode frequencies.  The evolutionary state (main sequence or shell
hydrogen burning) of Procyon A has the largest effect on the
predicted $p$-mode frequencies.  The $g$-modes show a greater
sensitivity to the various model parameters.  
\end{abstract}

\altaffiltext{1}{Hubble Fellow}

The Procyon binary system consists of an F5 IV-V primary and a
white-dwarf secondary in an 40.8 year orbit.  The F5 primary (Procyon
A) is a bright, nearby star with a well determined parallax and
astrometric mass.  As such, it presents a unique target for the study
of non-radial stellar oscillations.  Previous theoretical studies have
pointed out that the astrometric mass was incompatible with the mass
derived from stellar evolution calculations (Guenther \& Demarque
1993).  Procyon A was the target of a multi-site observing campaign
(10 -- 31 Jan/97) by the SONG project, whose goal was the detection of
stellar oscillations.  Motivated by these observations, as well as an
improved mass determination for this star, we have re-examined the
evolution and pulsation properties of Procyon A. The fundamental
properties of Procyon A are: $M = 1.50\pm 0.05\,\msun$ (Girard \ea\
1996), $\pi = 0.2832''\pm 0.0015$ (Girard \ea\ 1996), $F = (18.64\pm
0.87)\times 10^{-6}\, {\rm ergs\, cm^{-2}}\, s{^-1}$ (Smalley \&
Dworetsky 1995) and an angular diameter of $\phi = 5.51\pm 0.05\,$mas
(Mozurkewich \ea\ 1991) which imply $L= (7.22\pm 0.35)\,{\rm
L}_{\odot}$ and $R = (2.09\pm 0.02)\,{\rm R}_{\odot}$.

Models for Procyon A were calculated using the Yale stellar evolution
code, in its non-rotating configuration (Guenther \ea\ 1992).  These
models included the latest OPAL opacities (Iglesias \& Rogers 1996)
and the OPAL equation of state (Rogers \ea\ 1996). The diffusion
coefficients are from Thoul \ea\ 1994.  Most of the models only
included helium diffusion.  A single model which included both helium
and heavy element diffusion (treated as a single mean heavy element,
$Z$) was evolved to study the effects of heavy element diffusion on
Procyon A.  The models included the effects of a wind mass loss in the
diffusion equations.  A wind velocity of $v_w = -\dot{\rm M}/(4\pi
\rho r^2)$ was included in the diffusion equations, and a solar mass
loss rate $\dot{\rm M} = 2\times 10^{-14}\,\msun/{\rm yr}$ was
assumed.  This wind velocity was large enough to effectively suppress
the diffusion in the outer layers of the model.  One model was
calculated without wind loss.

The models for Procyon A were evolved from the ZAMS until they reached
the observed radius.  In an iterative procedure, the helium
abundance was adjusted in the ZAMS model and the model was re-evolved,
until the model matched the observed radius and luminosity of Procyon
A.  The starting helium abundance was constrained such that $1.5 <
\Delta Y/\Delta Z < 4$. In addition, the final surface abundance of $Z/X$ was
constrained to be within 0.1 dex of the solar value $Z/X = 0.0245$
(Grevesse \& Noels 1993), to satisfy the observed constraint that the
surface abundances of Procyon A are near solar.  Models
which met all of the above criteria had their pulsation frequencies
calculated using Guenther's nonradial nonadiabatic stellar pulsation
program (Guenther 1994).

In total, 8 different calibrated models of Procyon A were calculated
and pulsed (Table 1).  The standard model (line 1 in Table 1) uses the
best estimate for the mass ($1.5\,\msun$), does not include heavy
element diffusion or overshoot at the convective boundaries, and uses
our calibrated solar value for the mass fraction of the heavy elements
($Z$).  The other 7 models involved changing a single parameter from
the standard model: (a) low heavy element abundance (low $Z$); (b)
heavy element diffusion ($Z$ diff); (c) convective core overshoot of
0.1 pressure scale heights (overshoot); (d) evolution to a somewhat
lower luminosity (low L); (e) higher mass (high M) (f) lower mass
(subgiant); and (g) no wind loss. All but the subgiant model are in
the main sequence phase of evolution.  The lower mass model is in the
hydrogen shell burning phase of evolution.  
Table 1 includes the characteristic frequency spacing $\Delta$ of the
$p$-modes, which is approximately equal to the average frequency
separation between adjacent in $n$ $p$-modes, likely to be the first
quantity determined by stellar seismology.  Table 1 also includes the
characteristic period spacing $\Pi$ of the $g$-modes.
The frequencies and characteristic frequency spacings depend
on the radius of the star.  To first order, $\delta {\rm R_*/R_*} \simeq
\delta \nu/\nu$.  The radius of Procyon A is known to within 1\% .  
Thus, the calculated pulsation frequencies are uncertain at the
1\% level due to the error in the radius.  In addition,
uncertainties in the modeling of the superadiabatic layer (in both the
evolution and pulsation calculations) leads to an estimated
uncertainty of $\sim \pm 0.5\%$ in the $p$-mode frequency calculations
(see Guenther \& Demarque 1996).  The total error
associated with the calculated pulsation frequencies shown in Table 1
is approximately $\sim 1.5\%$.  Thus, differences in $\Delta$ greater
than $\sim 1\mu$Hz are significant.  From Table 1 we see that only the
subgiant model has a $\Delta$ which is significantly different from the
others.  This suggests that the detection of the average frequency
separation between adjacent in $n$ $p$-modes will be able to determine
the evolutionary status of Procyon A (or, indicate that the models are
in error).

\begin{table}
\caption{Model Characteristics}
\begin{center}\scriptsize
\begin{tabular}{lllcllllll}
\hline\hline
&
\multicolumn{1}{c}{Mass}&
&&&
\multicolumn{1}{c}{${\rm M_{core}}$}&
\multicolumn{1}{c}{${\rm M_{scz}}$}&
\multicolumn{1}{c}{$\log$}&
\multicolumn{1}{c}{$\Delta$}&
\multicolumn{1}{c}{$\Pi$}\\

\multicolumn{1}{c}{Model}&
\multicolumn{1}{c}{(${\rm M}_{\odot}$)}&
\multicolumn{1}{c}{$Z$}&
\multicolumn{1}{c}{$\Delta Y/\Delta Z$}&
\multicolumn{1}{c}{$Z/X_{\rm env}$}&
\multicolumn{1}{c}{(${\rm M}_{\odot}$)}&
\multicolumn{1}{c}{(${\rm M}_{\odot}$)}&
\multicolumn{1}{c}{$({\rm L/L}_{\odot})$}&
\multicolumn{1}{c}{($\mu$Hz)}&
\multicolumn{1}{c}{($\mu$Hz)}\\
\hline
Standard & 1.50 &  0.018 & 3.03  & 0.0253 & 0.118 & 
1.17E--4 & 0.85868  & 54.70 & 60.58\\

low Z    & 1.50 &  0.015 & 2.02   & 0.0206 &  0.107 & 
1.15E--4 & 0.85860 & 54.70 & 56.35 \\

Z diff   & 1.50 &  0.018 & 3.01  & 0.0253 & 0.118 & 
1.19E--4 & 0.85847 & 54.70 & 60.70 \\

overshoot& 1.50 &  0.018 & 2.53  & 0.0248 &  0.133 & 
1.04E--4 & 0.85848 & 54.73 & 64.77 \\

low L    & 1.50 &  0.018 & 2.47  & 0.0243 &  0.113 & 
2.31E--4 & 0.83747 & 55.02 & 59.35 \\

high M   & 1.54 &  0.018 & 1.90  & 0.0247 &  0.121 &
1.07E--4 & 0.85854 & 55.47 & 61.65 \\

subgiant & 1.40 &  0.018 & 2.94  & 0.0239 &  0.000 & 
5.62E--5 & 0.85838 & 52.91 & 16.87 \\

no wind  & 1.50 &  0.018 & 3.12  & 0.0196 &  0.122 & 
1.36E--5 & 0.85852 & 53.66 & 62.02 \\

\hline\hline
\end{tabular}
\end{center}
\label{tab1}
\end{table}

The characteristic period spacing
($\Pi$) of the $g$-modes shows a much larger variation than $\Delta$
(Table 1).  Once again, due to errors in the models and radii of
Procyon A, only differences in $\Pi$ greater than $\sim 1\mu$Hz are
significant.  In this context, it is clear that the detection of
$g$-modes in Procyon A would allow one to (a) determine its
evolutionary status; and (b) determine if appreciable overshoot 
($\ga 0.05\,{\rm H_p}$) is occurring at the edge of the convective core
in Procyon A, and/or  provide an estimate of the interior metallicity
of Procyon A.

\acknowledgments
BC was supported for this work by NASA through Hubble
Fellowship grant number HF--01080.01--96A awarded by the Space
Telescope Science Institute, which is operated by the Association of
Universities for Research in Astronomy, Inc., for NASA under contract
NAS 5--26555.

\end{document}